\documentclass[sigconf]{acmart}

\renewcommand\footnotetextcopyrightpermission[1]{}
\usepackage{booktabs} 
\usepackage{flexisym}
\usepackage{bbm}
\usepackage{balance} 
\usepackage{multirow}
\usepackage{float}

\usepackage{subfig}
\newcommand{\mathds}[1]{\text{\usefont{U}{dsrom}{m}{n}#1}}

\copyrightyear{2019}
\acmYear{2019} 
\setcopyright{iw3c2w3}
\acmConference[WWW '19 Companion]{Companion Proceedings of the 2019 World Wide Web Conference}{May 13--17, 2019}{San Francisco, CA, USA}
\acmBooktitle{Companion Proceedings of the 2019 World Wide Web Conference (WWW '19 Companion), May 13--17, 2019, San Francisco, CA, USA}
\acmPrice{}
\acmDOI{10.1145/3308560.3314201}
\acmISBN{978-1-4503-6675-5/19/05}

\begin{document}
\fancyhead{}

\title{Incorporating System-Level Objectives into Recommender Systems}

\author{Himan Abdollahpouri}
 \affiliation{Department of Information Science\\
  University of Colorado Boulder 
}
\email{himan.abdollahpouri@colorado.edu}

\begin{abstract}
One of the most essential parts of any recommender system is personalization-- how acceptable the recommendations are from the user's perspective. However, in many real-world applications, there are other stakeholders whose needs and interests should be taken into account. In this work, we define the problem of multistakeholder recommendation and we focus on finding algorithms for a special case where the recommender system itself is also a stakeholder. In addition, we will explore the idea of incremental incorporation of system-level objectives into recommender systems over time to tackle the existing problems in the optimization techniques which only look for optimizing the individual users' lists. 
\end{abstract}



%
%

\maketitle

\section{Problem}
Recommender systems provide personalized information access, supporting e-commerce, social media, news, and other applications where the volume of content would otherwise be overwhelming. They have become indispensable features of the Internet age, found in systems of many kinds. Even search engines, the most fundamental web applications, have become increasingly personalized in their provision of results, to the extent that they can also be considered recommender systems. 

One of the defining characteristics of recommender systems is personalization. Recommender systems are typically evaluated on their ability to provide items that satisfy the needs and interests of the end user. Such focus is entirely appropriate--Users would not make use of recommender systems if they believed such systems were not providing items that matched their interests. Still, it is also clear that, in many recommendation domains, the user for whom recommendations are generated is not the only stakeholder in the recommendation outcome. Other users, the providers of products, and even the system's own objectives may need to be considered. Fairness and balance are important examples of system-level objectives, and these social-welfare oriented goals may at times run counter to individual preferences. Sole focus on the end user hampers developers' ability to incorporate such objectives into recommendation algorithms and system designs. 

The integration of the perspectives of multiple parties into recommendation generation and evaluation is the underlying goal of the new sub-field of \textit{multistakeholder recommendation}~\cite{abdollahpouri_recommender_2017,soappaper,burke_educational_2016,nguyen2017multi}. The goal of a recommender system in a multistakeholder environment is, therefore, to generate recommendations taking all the stakeholder's needs and preferences into account.

\subsection{Multistakeholder Recommendation}
\label{sec:MS_recommendation}
In many recommender systems, there could exist a variety of different stakeholders. However, we can usually see three main stakeholders \cite{abdollahpouri2019beyond}:
 
\begin{description}
    \item[Consumers \textit{C}:] The consumers are those who receive the recommendations. They are the individuals whose choice or search problems bring them to the platform, and who expect recommendations to satisfy those needs.
    \item [Providers \textit{P}:] The providers are those entities that supply or otherwise stand behind the recommended objects and possibly gain utility from the consumer's choice. 
    \item [System \textit{S}:] The final category is the platform itself, which has created the recommender system in order to match consumers with providers and has some means of gaining benefit from successfully doing so. The platform may be a retailer, e-commerce site, broker, or other venue where consumers seek recommendations.
\end{description}

Typically, the system will attempt to satisfy the needs of at least one type of stakeholder by offering recommendations tailored (at least in part) to their preferences. When the stakeholder is the consumer, this is the most common personalized recommendation scenario. When there is personalization for both the consumer and the provider, we have a reciprocal recommendation.

In many real-world contexts, the system may gain some utility when recommending items, which could be in the form of a simple aggregate of the other stakeholders' utilities. In many e-commerce settings, the system will get a commission for each sale, and such benefits can be considered together with personalization~\cite{nguyen2017multi}. 

Alternatively, the system may seek to tailor outcomes specifically to achieve particular objectives. For example, an educational site may view the recommendation of learning activities as a curricular decision and seek to have its recommendations fit a model of student growth and development. Its utility may, therefore, be more complex than a simple aggregation of those of the other stakeholders.
One way to think about a multistakeholder recommendation is to look at it as a two-stakeholder system: the user and the system. In other words, in many applications, it is possible to aggregate the utilities of other stakeholders within the system utility and therefore optimizing for the system utility would, indirectly, optimize for their utility as well. Therefore, in this work, we focus on a scenario where we have two main stakeholders, the end-user and the system. 

In contrast to a single-stakeholder perspective which has been the case, for the most part, in recommender systems research, in this work we are interested in finding algorithms and mechanisms to incorporate system-level objectives into the recommendations. 

That being said, there are several challenges that need to be investigated:

\begin{itemize}
    \item How can the system-level preferences be incorporated into the recommendation generation?
    \item What are the different types of system-level preferences and how does the solution for incorporating each of them differs from the others?
    \item What is the right balance or trade-off between user preferences and system preferences? Does this balance vary depending on the domain?
    \item How can the system learn from the optimization in previous steps to do a better job at a particular time?
\end{itemize}

\section{State of the art}
There is a large body of recent work on incorporating diversity, novelty, long tail promotion and other metrics as additional objectives for recommendation generation and evaluation. See, for example, \cite{abdollahpouri2017controlling,diversityziegler}. There is also a growing body of work on combining multiple objectives using constraint optimization techniques, including linear programming. See, for example, \cite{jambor_optimizing_2010,jiang_optimization_2012,}. These techniques provide a way to limit the expected loss on one metric (typically accuracy) while optimizing for another, such as diversity. The complexity of these approaches increases exponentially as more constraints are considered, making them a poor fit for the general multistakeholder case. Also, for the most part, multiobjective recommendation research concentrates on maximizing multiple objectives for a single stakeholder, the end user.

\section{Proposed approach}
In this work, we will focus on one specific case of multistakeholder recommendation which is considering users and the system as two different stakeholders. That being said, we realized there is a substantial amount of work on using different types of non-accuracy metrics in recommender systems each of which used a different method to do the recommendation generation. For example, most of the work on recommendation list diversification, price optimization, fairness aware recommendation and so on use different algorithms to achieve the desired set of recommendation based on what evaluation metrics they use. In other words, there is a lack of understanding of what all these different problems have in common or from what aspects they differ. What is missing is the need for a general way of looking at extra objectives (i.e. system objectives) in recommender systems and to define all different types of such objectives and develop algorithms which can be used in problems with a similar type of objectives. What we propose is to have a standard framework that could be the home for all these different problems. To start, we need to categorize different types of system-level objectives. One main categorization in terms of the type of optimization that could be used to solve such problems is based on the scope of the system objective. Therefore, we divide every system-level objective into two main groups:

\begin{description}
    \item[Local objectives:] in this group we have objectives that could be optimized within each user's recommendation list. In other words, there is no need to look at the entire user population in order to do the optimization. List diversification and novelty (from the user's perspective) belongs to this group.
    \item[Global objectives:] The second group is the type of objectives that need to be taken care of not only based on the recommendations within each user's list but also the recommendation lists of other users as well. Objectives like long-tail promotion and fair exposure belong to this category.  
\end{description}

In addition to this two classes of objectives, we can also define several other types of objectives based on whether the system's focus is on 1) items, 2) users or maybe 3) both at the same time. We define different types of system objectives with respect to which entity the system is focusing on as follow:

\begin{description}
    \item[User-item related:] The system could have a certain goal regarding which user should get what recommendation. This is more like a parental control which means, in addition to what users like, the system also has to decide whether that item suits the user or not. An example would be in educational recommendation where the system might not want to recommend a program to a user even though she likes it or, recommend a certain program even though she may not like it the most compared to other programs. 
    \item[user related:] When the system wants the users who receive a certain item (items) as the recommendation to have certain properties. For example, in an educational recommender system, the goal might be to make sure the users who receive the recommendation are \%50 male and \%50 female. Many of the fairness-aware recommendation problems fall in this category.
    \item[item related:] This is when the system has a certain goal regarding a certain group of items. For example, long-tail recommendation, which is a well-known problem, is when the system wants to promote long-tail items more often. 
 
 \end{description}

\section{Results and the Progress so far}
As a very specific case of system-level objective, we have worked on the problem of long-tail promotion (i.e. controlling popularity bias) in recommender systems. 
Popularity bias refers to the situation in which there is an over-concentration of the recommendations of certain items, the popular ones. Controlling popularity bias is important for many recommender systems as it affects the fairness of the platform and it also improves the novelty of the recommendations. For more information on popularity bias see ~\cite{abdollahpouri2017controlling,himanaies}main.\\
We wanted to go a bit deep in this problem and find different types of solutions for it to make a better sense of this type of objective and to be able to generalize for similar problems. Our aim is, however, to find a general class of solutions for other types of system objectives for the next steps of this research.

In order to control the effect of popularity bias in recommender systems we have used two approaches: 
One way is to modify the underlying recommendation algorithm such that we take the popularity of items into account in addition to the relevance of those items for generating the predicted ratings. Therefore, the final list is both personalized and also controlled for popularity bias. 

Another approach is to keep the existing algorithm untouched but add an extra level of item filtering on the predicted ratings. In this method, the algorithm tries to re-rank the items based on both their accuracy (predicted rating) and popularity. 

\subsection{Technique 1: Model based Approach: Fairness-aware Regularization}
\noindent

In this approach, we have explored the use of regularization to control the popularity bias of a recommender system. We start with an optimization objective of the form:

\begin{equation}
     \min\limits_{P,Q} acc(P,Q)+ \lambda  reg(P,Q)
\end{equation}
where $acc(.)$ is the accuracy objective, $reg(.)$ is a regularization term, and $\lambda$ is a coefficient for controlling the effect of regularizer.

Our goal, therefore, is to identify a regularization component of the objective function that will be minimized when the distribution of recommendations is fair in terms of popular and non-popular items.  

We define two sets of items ($\Gamma'$) and ($\Gamma$) corresponding to the popular and non-popular items, and define a co-membership matrix $D$, over these sets. For any pair of items $i$ and $j$, $d(i,j)=1$ if $i$ and $j$ are in the same set and $0$ otherwise.
We define intra-list binary unfairness (ILBU) as the average value of $d(i,j)$ across all pairs of items $i,j$. 

\begin{equation}
     ILBU(L_{u})=\frac{1}{N(N-1)}\sum_{i,j \in L_{u}} d(i,j)
\end{equation}
where $N$ is the number of items in the recommendation list. The fairest list is one that contains equal numbers of items from each set, which can be easily seen.

For more details on how the optimization is solved please refer to \cite{abdollahpouri2017controlling}.

\subsection{Technique 2: Post processing Re-ranking}
\subsubsection{xQuAD}
For the second approach, we used a post-processing re-ranking technique to diversify the recommendation lists in terms of popular and unpopular items. We have used EXplicit Query
Aspect Diversification (xQuAD) \cite{santos2010exploiting} which explicitly accounts for the various aspects associated with an under-specified query. In the context of search engines, items are selected iteratively by estimating how well a given document satisfies an uncovered aspect. We build on the \textit{xQuAD} model to control popularity bias in recommendation
algorithms. We assume that for a given user $u$, a ranked recommendation list $R$ has already been generated by a base recommendation algorithm. The task of the modified xQuAD method is to produce a new re-ranked list $S$ ($|S|<|R|$) that manages popularity bias while still being accurate. 

The new list is built iteratively according to the following criterion:
\begin{equation}\label{eq:1}
    P(v|u)+\lambda P(v,S'|u)
\end{equation}
where $P(v|u)$ is the likelihood of user $u \in U$ being interested in
item $v \in V$, independent of the items on the list so far as predicted by the base recommender. The second term $P(v, S'|u)$ denotes the likelihood of user u being interested in an item $v$ as an item not in the currently generated list $S$.

Intuitively, the first term
incorporates accuracy while the second term promotes
diversity between two different categories of items (i.e. short head and long tail). The parameter $\lambda$ controls how strongly controlling popularity bias is weighted in general. The item
that scores most highly under the equation \ref{eq:1} is added to the
output list $S$ and the process is repeated until $S$ has achieved the
desired length. 

For more details on these two techniques refer to \cite{abdollahpouri2017controlling} and \cite{flairs2019}. 

we tested our proposed algorithms on two public datasets. We show the result for one dataset due to lack of space but a, more or less, a similar result could be seen for the other dataset. The dataset for which you can see the result in figure 1, is the Epinions dataset which is gathered from a consumers opinion site where users can review items \cite{massa2007trust}. This dataset has the total number of 664,824 ratings given by 40,163 users to 139,736 items. We split the items in both datasets into two categories: long-tail ($\Gamma$) and short head ($\Gamma$')in a way that short head items correspond to \%80 of the ratings while long-tail items have the rest of the \%20 of the ratings. 

\subsection{Evaluation}
The experiments compare four algorithms. Since we are concerned with ranking performance, we chose as our baseline algorithm RankALS, a pair-wise learning-to-rank algorithm. We used the output from RankALS as input for the re-ranking technique described above. Note that, we developed two versions of the re-ranking algorithm:  Binary xQuAD and Smooth xQuAD, marked \textit{Binary} and \textit{Smooth} in the figures. The difference between the two as you can read in more details in \cite{flairs2019} is in how they check for whether an item category (popular or non-popular) is already covered by the list or not. The binary version only returns 0 (not covered) or 1 (covered) while the smooth version returns what percentage of the list is covered with each item category. Also, the model-based approach is shown by \textit{LT_Reg} in the figure. We compute lists of length 100 from RankALS and pass these to the re-ranking algorithms to compute the final list of 10 recommendations for each user.
In order to evaluate the effectiveness of algorithms in mitigating popularity bias, we used four different metrics each of which measures a certain aspect of long-tail promotion.

\textbf{Average Recommendation Popularity (ARP)}: This measure from \cite{yin2012challenging} calculates the average popularity of the recommended items in each list. For any given recommended item in the list, we measure the average number of ratings for those items. More formally:
\begin{equation}\label{eq:arp}
     ARP=\frac{1}{|U_{t}|}\sum_{u \in U_{t}} \frac{\sum_{i \in L_u}\phi(i)}{|L_u|}
\end{equation}
where $\phi(i)$ is the number of times item $i$ has been rated in the training set. $L_u$ is the recommended list of items for user $u$ and $|U_{t}|$ is the number of users in the test set.

\textbf{Average Percentage of Long Tail Items (APLT)}: As used in \cite{abdollahpouri2017controlling}, this metric measures the average percentage of long tail items in the recommended lists and it is defined as follows:
\begin{equation}\label{eq:aplt}
     APLT=\frac{1}{|U_{t}|}\sum_{u \in U_{t}} \frac{|\{i, i \in (L_u \cap \Gamma)\} |}{|L_u|}
\end{equation}

This measure gives us the average percentage of items in users' recommendation lists that belong to the long tail set.

\textbf{Average Coverage of Long Tail items (ACLT)}: This is the metric we introduced in \cite{flairs2019} which measures how much exposure long-tail items get in the entire recommendations. One problem with $APLT$ is that it could be high even if all users get the same set of long tail items. $ACLT$ measures what fraction of the long-tail items the recommender has covered: 
\begin{equation}\label{eq:aclt}
    ACLT=\frac{1}{|U_{t}|}\sum_{u \in U_{t}} \sum_{i \in L_u} \mathds{1}({i \in \Gamma})
\end{equation}
where $\mathds{1}({i \in \Gamma})$ is an indicator function and it equals to 1 when i is in $\Gamma$. This function is related to the \textit{Aggregate Diversity} metric of \cite{adomavicius2012improving} but it looks only at the long-tail part of the item catalog.

In addition to the aforementioned long tail diversity metrics, we also evaluate the accuracy of the ranking algorithms in order to examine the diversity-accuracy trade-offs. For this purpose we use the standard \textit{Normalized Discounted cumulative Gain} (\textit{NDCG}) measure of ranking accuracy. 

Figure~\ref{fig:ep-results1} shows the results for the Epinions dataset across the different algorithms using a range of values for $\lambda$. (Note that the LT-Reg algorithm uses the parameter $\lambda$ to control the weight placed on the long-tail regularization term.) All results are averages from five-fold cross-validation using a \%80 -\%20 split for train and test, respectively. As expected, the diversity scores improve for all algorithms, with some loss of ranking accuracy. Differences between the algorithms are evident, however. The exposure metric (ACLT) plot shows that the two re-ranking algorithms, and especially the Smooth version, are doing a much better job of exposing items across the long-tail inventory than the regularization method. The ranking accuracy shows that, as expected, the Binary version does slightly better as it performs minimal adjustment to the ranked lists. LT-Reg is not as effective at promoting long-tail items, either by the list-wise APLT measure or by the catalog-wise ACLT.

\begin{figure*}[t]
    \centering
    \includegraphics[width=6in]{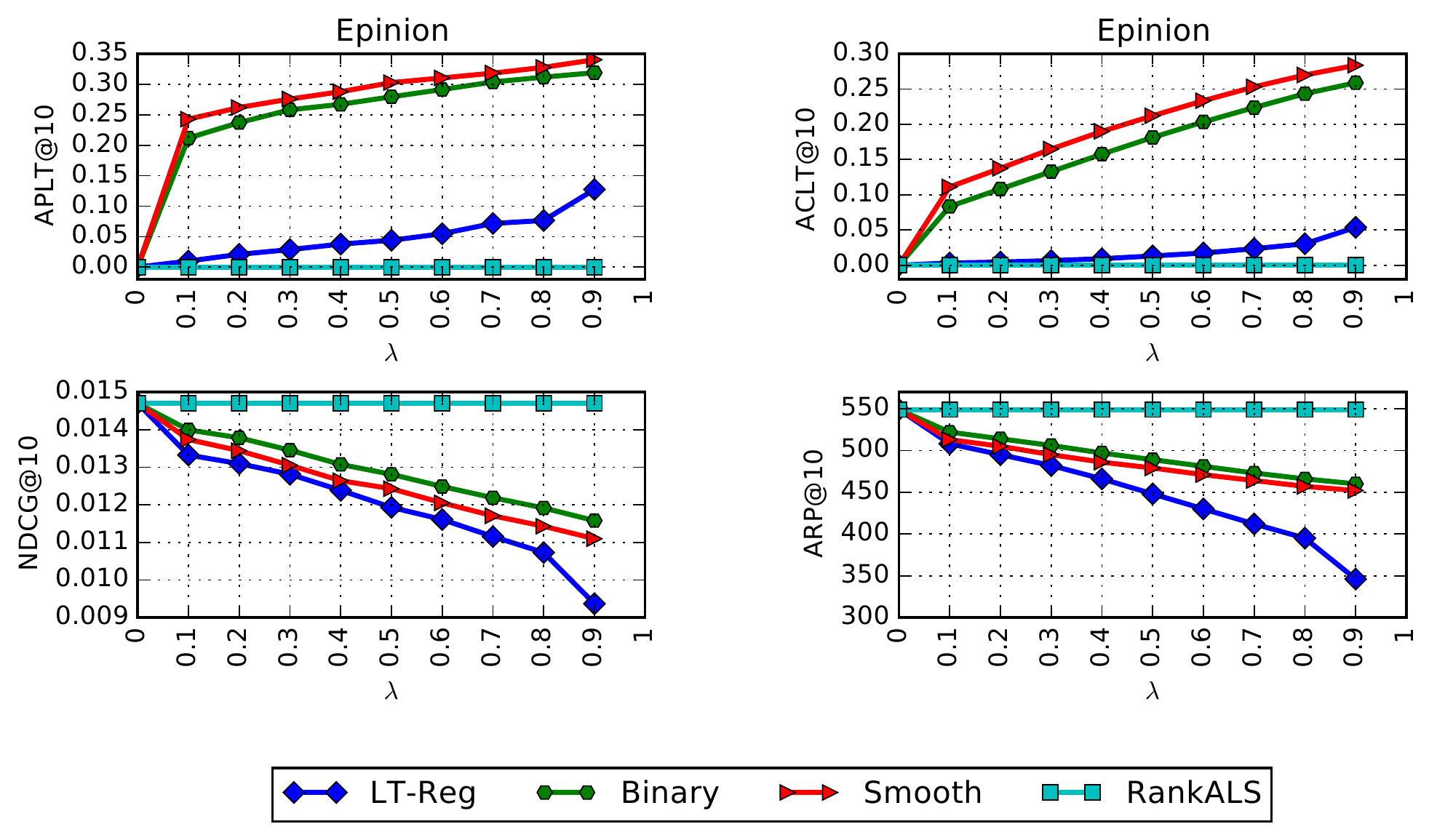}
    \caption{Results for the Epinions dataset}
    \label{fig:ep-results1}
\end{figure*}

\section{Conclusions and future work}

\begin{figure*}[t]
    \centering
    \includegraphics[width=4in]{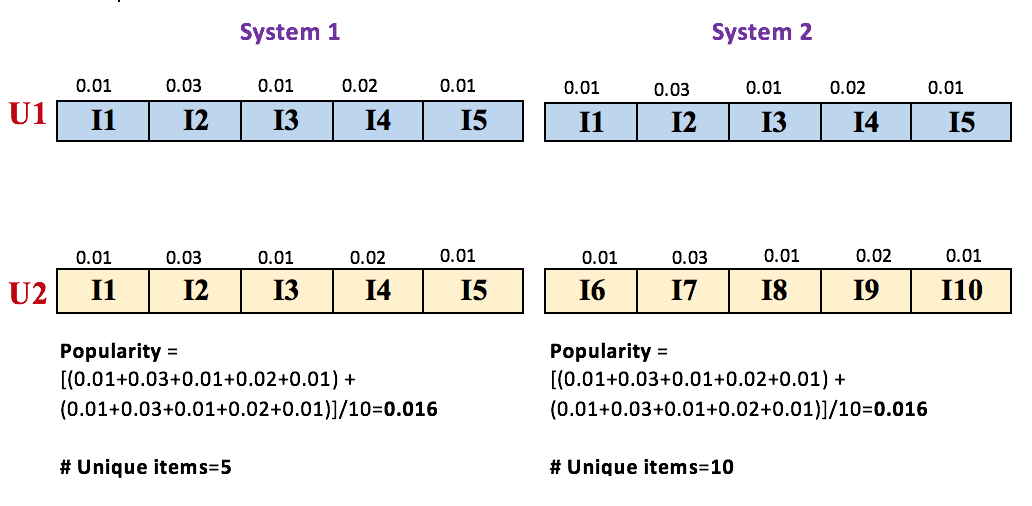}
    \caption{The recommendation lists of length 5 to different users U1 and U2 generated by two different recommender systems. Both systems have similar sum of popularity, but system 2 has covered 10 items while system 1 has only covered 5.}
    \label{fig:ep-results1}
\end{figure*}

Working on tackling popularity bias as an example of system-level objective, we learned that the system objective could be incorporated in three different ways which we only explored two of them: 1) Model-based multi-objective optimization, 2) re-ranking approach, and 3) data manipulation which we have not yet worked on. The work we have done so far, might not be, just by itself, a great contribution to the field as there are similar approaches for tackling popularity bias but it inspired us to look at the problem in a more general way as popularity bias is not the only objective that a system could have. For example, \textit{user-level list diversity}, \textit{aggregate diversity}, \textit{profit maximization}, \textit{fairness-aware recommendation}, to name just a few, all could be considered as system-level objectives which should be incorporated into the recommendation process. So one future work is to find a general framework by which we can incorporate any type of system-level objectives into the recommender systems. 

Another important observation we made which is also a possibility for future work is that almost all the works on long-tail promotion have done the optimization on individual's recommendation lists hoping this will lead to an overall optimization within the entire user base. However, as we saw in our experiments (ACLT vs ARP), a system could have a low overall sum of popularity for the recommendations but might not have done a good job of promoting enough long-tail items. The reason is optimizing the individual's recommendation list (local optimization) does not guaranty an overall optimization (global optimization) and, therefore, a more dynamic and temporal optimization technique is needed. Figure 2 shows a simple scenario where two systems have generated recommendations for two users. The popularity values for each item is shown on the top of each recommended item. A popularity value of 0.01 means the item has been rated by \%1 of the users.  As you can see, both systems have the same sum of popularity but one has covered 10 items while the other has only covered 5. This example emphasizes the need for better incorporation of system-level objectives into recommendations. One possibility to address this problem is to use incremental optimization \cite{sharp2007incremental} over time to balance system and user-level objectives. Figure 3 shows a schematic view of how the optimization could be done over time such that, at any given time T, the recommender system optimizes the objective function by also looking at what the outcome for optimization was at previous times. 

\begin{minipage}{\linewidth}
\makebox[\linewidth]{
  \includegraphics[keepaspectratio=true,scale=0.50]{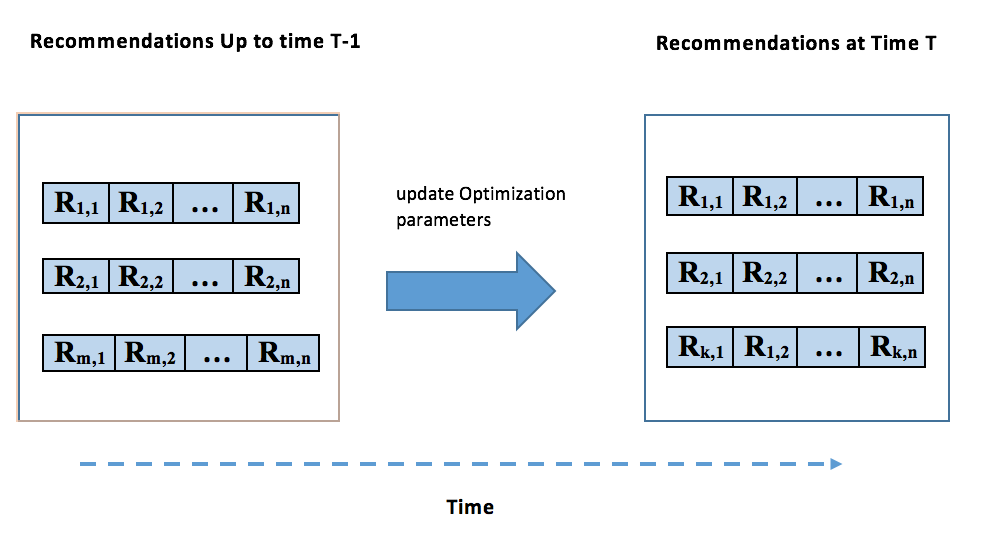}}
\captionof{figure}{Dynamic optimization of the recommendations over time}\label{visina8}
\end{minipage}

\bibliographystyle{ACM-Reference-Format}
\balance
\bibliography{main.bib}

\end{document}